\documentclass{aa}

\usepackage{graphicx}
\usepackage{txfonts}

\usepackage{natbib}

\usepackage{soul}
\usepackage{color}
\usepackage[colorlinks=true,citecolor=blue]{hyperref}
\begin{document}

   \title{The obliquity and atmosphere of the ultra-hot Jupiter TOI-1431b (MASCARA-5b):
  A misaligned orbit and no signs of atomic or molecular absorptions.  }
\titlerunning{The atmosphere of the ultra-hot Jupiter TOI-1431b (MASCARA-5b)}

   \author{M. Stangret\inst{1}\fnmsep\inst{2}, E. Pallé\inst{1}\fnmsep\inst{2}, N. Casasayas-Barris\inst{1}\fnmsep\inst{2}\fnmsep\inst{3}, M. Oshagh\inst{1}\fnmsep\inst{2}, A. Bello-Arufe \inst{4}, R. Luque\inst{1}\fnmsep\inst{2}, V. Nascimbeni\inst{5}, F. Yan\inst{6}, J. Orell-Miquel\inst{1}\fnmsep\inst{2}, D. Sicilia\inst{7}, L. Malavolta\inst{5}\fnmsep\inst{8}, B. C. Addison\inst{9}, L. A. Buchhave\inst{4}, A. S. Bonomo\inst{10}, F. Borsa\inst{11}, S. H. C. Cabot\inst{12}, M. Cecconi\inst{13},  D. A. Fischer\inst{12}, A. Harutyunyan\inst{13}, J. M. Mendonça\inst{4},  G. Nowak \inst{1}\fnmsep\inst{2}, H. Parviainen\inst{1}\fnmsep\inst{2}, A. Sozzetti\inst{10}, R. Tronsgaard \inst{4}}
   
   \authorrunning{M. Stangret et al.}

   \institute{Instituto de Astrofísica de Canarias, Vía Láctea s/n, 38205 La Laguna, Tenerife, Spain \\
             \email{mstangret@iac.es}          
             \and 
             Departamento de Astrofísica, Universidad de La Laguna, 38200 San Cristobal de La Laguna, Spain
             \and 
             Leiden Observatory, Leiden University, Postbus 9513, 2300 RA Leiden, The Netherlands
             \and
             National Space Institute, Technical University of Denmark, Elektrovej, DK-2800 Kgs. Lyngby, Denmark
             \and
             INAF – Osservatorio Astronomico di Padova, Vicolo dell’Osservatorio 5, I-35122, Padova, Italy
             \and
             Institut f\"ur Astrophysik, Georg-August-Universit\"at, Friedrich-Hund-Platz 1, 37077 G\"ottingen, Germany
             \and
             INAF – Osservatorio Astrofisico di Catania, Via Santa Sofia 78, 95123 Catania, Italy
             \and
             Department of Physics and Astronomy, Università degli Studi di Padova,
            Vicolo dell’Osservatorio 3, I-35122, Padova, Italy
            \and
            University of Southern Queensland, Centre for Astrophysics,
            USQ Toowoomba, West Street, QLD 4350 Australia
             \and
             INAF – Osservatorio Astrofisico di Torino, via Osservatorio 20, 10025 Pino Torinese, Italy
             \and
             INAF – Osservatorio Astronomico di Brera, Via E. Bianchi 46, 23807 Merate, Italy
              \and
              Yale University, 52 Hillhouse Avenue, New Haven, CT 06511, USA
             \and
             Fundación Galileo Galilei - INAF, Rambla José Ana Fernandez Pérez 7, E-38712, Breña Baja, TF - Spain
             }

   \date{Received XXX; accepted XXX}

 
  \abstract
{ Ultra-hot Jupiters are defined as giant planets with equilibrium temperatures larger than 2000 K. Most of them are found orbiting bright A-F type stars, making them extremely suitable objects to study their atmospheres using high-resolution spectroscopy. Recent studies show a variety of atoms and molecules detected in the atmospheres of this type of planets. Here we present our analysis of the newly discovered ultra-hot Jupiter TOI-1431~b/MASCARA-5~b, using two transit observations with the HARPS-N spectrograph and one transit observation with the EXPRES spectrograph. Analysis of the Rossiter-McLaughlin effect shows that the planet is in a polar orbit, with a projected obliquity $ \lambda = -155^{+20}_{-10}$ degrees. Combining the nights and applying both cross-correlation methods and transmission spectroscopy, we find no evidences of \ion{Ca}{I}, \ion{Fe}{I,} \ion{Fe}{II}, \ion{Mg}{I}, \ion{Na}{I}, \ion{V}{I}, TiO, VO or H$\alpha$ in the atmosphere of the planet. Our most likely explanation for the lack of atmospheric features is the large surface gravity of the planet.
 }


   \maketitle
 
%

\section{Introduction}

Ultra-hot Jupiters (hereafter UHJ) are giant planets with equilibrium temperatures higher than 2000 K \citep{helling_wasp_18}, caused by their short orbital distance and the strong irradiation from their host star. Theoretical and observational studies show that due to thermal dissociation one does not expect \ion{H$_2$O}{}in their dayside atmospheres \citep{Parmentier2018}, as is the case of hot Jupiters. On the other hand, the elevated temperatures cause many atomic elements to be found in the ionized state. Combined with the fact that most of UHJs are found around bright A type stars, this makes UHJs perfect laboratories to detect and study their atmospheres. Thanks to the different Doppler velocities of the planets and their host stars, as well as the Earth, we are able to study the planetary atmospheres using high-resolution spectroscopy \citep{2018arXiv180604617B, Snellen2010} from ground-based spectrographs.

Recent studies show variety of atoms and molecules detected in the atmospheres of ultra-hot planets. In the atmosphere of the hottest planet known to date, KELT-9b \citep{gaudi_kelt_9b}, \ion{Ca}{II}, \ion{Cr}{II}, \ion{Fe}{I}, \ion{Fe}{II}, \ion{Mg}{II}, the Mg triplet, \ion{H}{}, \ion{Na}{I}, \ion{Sc}{II}, \ion{Ti}{II} and \ion{Y}{II}, as well as evidence of \ion{Ca}{I}, \ion{Cr}{I}, \ion{Co}{I} and \ion{Sr}{II} were detected \citep{Hoeijmakers_2018_kelt9, Hoeijmakers_2019_kelt9, Cauley_2019_kelt9, yan-2019-calcium-kelt9-wasp33, turner_2020_kelt9-ca, pino_2020_kelt9_FeI}. KELT-9b also possesses an extended hydrogen atmosphere detected in H{$\alpha$} absorption \citep{YanKELT9}. Several other UHJs have been explored so far and show the detection of one or several of the above mentioned species: WASP-33b \citep{yan-2019-calcium-kelt9-wasp33, nugroho_2020_wasp33_FeI, Yan_2021_wasp33}, WASP-12b \citep{Jensen_2018_wasp12}, WASP-76b \citep{seidel-2019-wasp-76, Ehrenreich_wasp76}, WASP-189b \citep{Yan_2020_wasp-189}, WASP-121b \citep{2020Cabot_wasp-121, 2020BenYami_wasp121,2020arXiv200106430G, 2020HoeijmakersHEARTS} and MASCARA-2b/KELT-20b \citep{Casasayas2018, Casasayas2019, Stangret_2020_MASCARA-2, Nugroho2020_KELT20, Hoeijmakers_mascara2}.

TOI-1431b, also known as MASCARA-5b, is a newly discovered ultra-hot Jupiter with an equilibrium temperature of 2181 K, orbiting bright A star (V=8.049 mag) in 2.6502409 days (Addison et al., 2021 submitted). The planet was originally discovered using the ground-based MASCARA survey. More recently, the NASA TESS mission \citep{Ricker2015} also observed the transits and alerted it as TOI-1431b. MASCARA-5b has been confirmed as a planet by Addison et al. 2021 (submitted) using ground-based facilities.

Here we present the analysis of three transit observations of TOI-1431b using high-resolution spectroscopic observations, which are described in section 2, in order to retrieve the system's architecture as well as to explore the planetary atmosphere. In section 3 we measure the geometry of the system by analysing the Rossiter–McLaughlin effect and in sections 4 and 5 we explore the composition of the atmosphere using cross-correlation and transmission spectroscopy, respectively.


\section{Observations}

We observed two full transits of TOI-1431b during the nights of 31 May 2020\footnote{Collected during ITP\,19-1 program (PI: Pall\'e).} (hereafter Night 1) and 23 July 2020\footnote{Collected during GAPS2 Long-Term program (PI: G. Micela).} (hereafter Night 2) using the HARPS-N spectrograph \citep{Cosentino_2012_harpsn} mounted at the 3.58 m Telescopio Nazionale Galileo (TNG) at Observatorio del Roque de los Muchachos (ORM) in La Palma, Spain. During the first night we took 60 exposures of 300 s, resulting in 33 out-of-transit and 27 in-transit spectra (covering the range $\phi$=-0.022 to +0.065; where $\phi$ is planet orbital phase), with an average signal-to-noise ratio (S/N) of 97.4. During the second night we took 61 exposures of 300 s, resulting in 31 out-of-transit and 30 in-transit spectra ($\phi$=-0.039 to +0.046), with an average S/N of 66.

TOI-1431b was also studied using the EXtreme PREcision Spectrograph (EXPRES), an optical high resolution ($R \sim 140,000$) fiber-fed echelle instrument commissioned at the 4.3 m Lowell Discovery Telescope (LDT, \citealt{2012Levine}), at Lowell Observatory, covering the wavelength range 380 - 680 nm. Although EXPRES was designed with the primary goal of detecting Earth-like exoplanets around Sun-like stars \citep{2016Jurgenson}, \cite{Hoeijmakers_mascara2} showed it can also be used for the study of exoplanet atmospheres.  A full transit of TOI-1431b was observed with EXPRES during the night of 28 July 2020. We obtained 49 exposures of 330 seconds ($\phi$=-0.045 to +0.038), resulting in 25 out-of-transit and 24 in-transit spectra. A summary log of all the observations is given in Table~\ref{table:log}.

The data from HARPS-N were reduced using the HARPS-North Data Reduction Software (DRS,  \citet{Cosentino_2012_harpsn, DRS2}), version 3.7, which allows us to extract the spectra order-by-order and use a daily calibration set to perform flat-field. In the final steps, a one-dimensional spectrum (380 nm - 690 nm in a step of 0.01 nm) is created by combining all the orders for each spectrum separately. The EXPRES data were reduced using the EXPRES pipeline described in \cite{2020Petersburg}, which performs telluric correction using SELENITE \citep{2019Leet}.

\begin{table*}
\caption{Summary of the transit observations.}             
\label{table:log}      
\centering          
\resizebox{\textwidth}{!}{\begin{tabular}{c  c c c c c c c c c  }     
\hline\hline       
Object  & Instrument & Date of observation & Start UT& End UT& Texp (s)& airmass\tablefootmark{a} & S/N@588nm\tablefootmark{a} & Nobs  \\ 
\hline     
    & &   &  &  &  &  &  & \\ [-0.7em] 
TOI-1431b  & HARPS-N & 2020-05-31 & 23:45 & 05:13 & 300 & 1.12-2.34 & 107-129 & 60\\  
TOI-1431b  & HARPS-N & 2020-07-23 & 22:39 & 04:04& 300 & 1.12-1.41 & 52-93 & 61\\ 
\hline  
TOI-1431b  & EXPRES & 2020-07-29   & 05:34&10:51 & 330 & 1.07-1.23 &-  & 49\\ 
\hline

\end{tabular}}
\tablefoot{\tablefoottext{a}{Minimum and maximum values during the night.}}
\end{table*}

\begin{table}
\small
\caption{Physical and orbital parameters of TOI-1431 adopted from Addison et al. 2021(submitted). Parameters marked with * were calculated in the current work.}             
\label{Tab:parameters}      
\centering                          
\begin{tabular}{l l l}        
\hline\hline                 
Description & Symbol & Value \\    
\hline                        
   Identifiers & - &  TOI-1431, HD 201033  \\[0.1em]       
   V-band magnitude & $m_V$ & $8.049 \pm 0.011 $ mag    \\[0.1em]
   Effective temperature & $T_{eff}$ &   $7690 ^{+400}_{-250}$ K  \\[0.1em] 
  Surface gravity  & $ \log{g}$ &  $ 4.148 ^{+0.043}_{-0.041}$     cgs \\[0.1em]
  Metallicity  & [Fe/H] & $ 0.43^{+0.20}_{-0.28}$    \\[0.1em]
  Stellar mass  & $M_{\star} $ &   $1.895^{+0.100}_{-0.077} $  M$_\odot$ \\[0.1em]
  Stellar radius  & $ R_{\star}$ &  $  1.923^{+0.068}_{-0.067}$ R$_\odot$  \\[0.1em]
  \hline
 Planet mass & $M_{p} $ &  $3.12^{+0.19}_{-0.18} $ M$_J$   \\[0.1em]
  Planet radius & $R_{p} $ &  $  1.546\pm0.063 $ R$_J$    \\[0.1em]
  Equilibrium temperature  & $T_{eq} $ &  $ 2181 \pm 95 $ K  \\[0.1em]
  \hline
  Right ascension  & $...$ & $ 21^h 04^m  48.89^s $    \\[0.1em]
  Declination  & $...$ & $ +55^{o} 35^{'} 16.88^{''}$    \\[0.1em]
 Mid Transit Time (BJD) & $T_{0}  $ & $2458739.17728 \pm 0.00011$   \\[0.1em]
  Period  & $ P$ & $  2.6502409 \pm 0.0000041$ days   \\[0.1em]
  Transit duration  &  $T_{14} $& $0.1047 \pm 0.0006$  days  \\[0.1em]
  Semi-major axis & $ a$ & $0.0465 \pm 0.0017$ AU   \\[0.1em]
  Inclination  & $i $ &  $80.30^{+0.18}_{-0.17}$ deg  \\[0.1em]
  Eccentricity  & $ e$ &  $0.0051{+0.016}_{-0.0039} $   \\[0.1em]
  Projected obliquity\tablefootmark{*}  & $ \lambda$ &  $-155^{+20}_{-10}$  deg  \\[0.1em]
Projected rotation speed\tablefootmark{*}    & $v \sin{i_\star}_{RM}  $ & $3.73^{+1.29}_{- 1.28}  $ km\,s$^{-1}$ \\[0.1em]  

Planetary RV  & $K_p$ & 188.15  km\,s$^{-1}$   \\
 semi-amplitude & &  \\[0.1em]
  Linear limb darkening   & $\epsilon $&  $0.65\pm0.20$ \\[0.1em]
  coefficient\tablefootmark{*}  & &  \\[0.1em]
   $A_{GP}$\tablefootmark{*}  & $...$&  $1.62^{+0.81}_{-0.50} $~${\rm  m~s^{-1}}$\\[0.1em]
 $\tau_{GP} $\tablefootmark{*}  & $...$&  $0.02^{+0.01}_{-0.00} $ days\\[0.1em]
\hline                                   
\end{tabular}
\end{table}

\section{Rossiter–McLaughlin analysis}
\label{sec:rm}

There are two main approaches to extract the radial velocities (RVs) during the transit of an exoplanet and obtain the Rossiter–McLaughlin (RM) signal. One approach relies on the template matching of the observed spectra \citep{Butler-96, Guillem_TERRA}, and the other one is based on a Gaussian fit to the cross-correlation function (CCF) of the observed spectra with a binary mask \citep{Pepe-02}. We calculated the RVs during two transits of TOI-1431b observed by HARPS-N from both template matching approach using SERVAL \citep{Zechmeister-18}, and CCF approach using DRS \citep{Cosentino_2012_harpsn}. For the template matching approach for each observing night the template was created from only the out-of-transit spectra. For the HARPS-N data, the RM signals extracted from CCF approach show a larger variation in the out of transits RV measurements, and also a higher number of outliers. This could be due to using an inadequate binary mask (G2 mask) for this type of host star. Thus, for the rest of the analysis, we decided to focus only on RM results from the template matching approach. For the transit obtained from EXPRES, the RVs were derived with the forward modeling from empirical stellar spectral templates, as described in \citet{2020Petersburg}, which is similar to the template matching approach.

\begin{figure}[h!]
	\centering
    \includegraphics[width=0.5\textwidth, height=7cm]{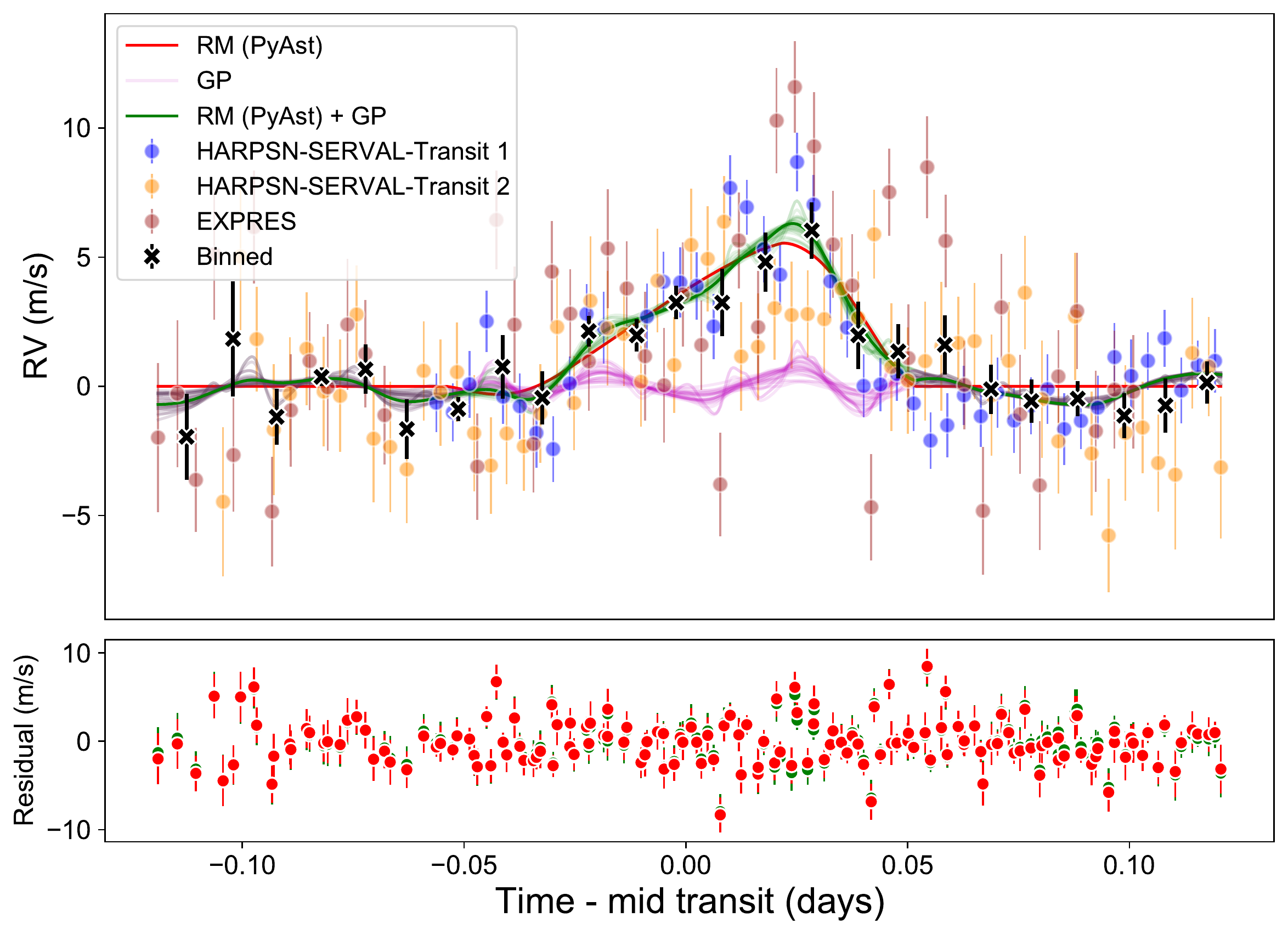}
      \caption{Radial velocity time series derived using the template matching approach for each of the three transits. Also shown is the the binned data from combining all three transits. The best fit model to the folded RM is shown using a \texttt{PyAstronomy} model incorporating a GP. The different components of the best fit model are plotted in different colors as marked in the legend. The residual between the best fit model using \texttt{PyAstronomy} and GP+\texttt{PyAstronomy} and observations are shown in red and green circles, respectively, in the bottom panel.}
  \label{fig:PyastronomyGP}
\end{figure}

Stellar noise can cause an offset and underlying slopes in the RV measurements in out-of-transit RM observations. This slope is in addition to the gravitationally induced RV variation generated by the orbiting planet. The activity-induced out-of-transit RV slope can differ from transit to transit due to variations in stellar activity over different nights \citep{Oshagh-18}. A conventional approach to eliminate this effect is to remove a linear trend considering only the out-of-transit RVs. Therefore, here we also removed a linear trend in RVs fitted to each individual night, and we analyse the combined and folded RM observations, as shown in Figure~\ref{fig:PyastronomyGP}.

To model the observed RM signal we use the prescription presented in \citet{Ohta-05}, which is optimized to retrieve the RM signal from template matching. This model is implemented in the \texttt{PyAstronomy} python package \citep{pya}. 

Gaussian process (GP) is a widespread framework for modeling correlated noise \citep{Rasmussen-06}, and its power and advantages in mitigating correlated noise in RV observations \citep[e.g.,][]{Haywood-14, Faria-16}, and also in photometric transit observations \citep[e.g.,][]{Aigrain-16, Serrano-18}, has been demonstrated widely. Our three RM observations show clear variation especially in the out of transit RV measurements, which could be due to either stellar noise (either granulation or active regions), an instrumental systematics, or telluric contamination. Therefore, we decided to perform two independent analysis, one without considering GP and one that incorporates GP to our RM modeling. We used the recent implementation of GP in \texttt{celerite} package \citep{Foreman-Mackey-17}, as some of the \texttt{celerite} kernels are well suited to describe different forms of correlated noise.

We consider the spin-orbit angle $\lambda$, projected stellar rotational velocity ($v \sin{i_\star}$), mid-transit time ($T_0$), and limb darkening coefficient as our mean model's free parameters. The rest of the parameters required in the mean model are fixed to their reported values in Table~\ref{Tab:parameters}. The posterior samples for our model were obtained through MCMC (Markov chain Monte Carlo) using emcee \citep{Foreman-Mackey-13}. The priors on $v \sin{i_\star}$ and $T_0$ are controlled by Gaussian priors centered on the reported value in Table~\ref{Tab:parameters} and with broader widths than the reported uncertainties, to allow better exploration of the parameter space. The prior on the linear limb-darkening coefficient were also constrained by Gaussian prior created using \texttt{LDTk} \citep{Parviainen-15}. The prior on spin-orbit angle is controlled by a uniform (uninformative) prior between -180 and +180 degrees. 

For the case of \texttt{PyAstronomy}+GP, we fit our RM observations considering the sum of a mean model and noise model. The mean model is the RM model of \texttt{PyAstronomy}, and the noise was modeled as a GP with Matern-3/2 covariance Kernel \citet{Palle2020}. This Kernel is well suited
to describe different forms of stellar noises. The prior on the GP's time scale parameter was controlled by a Gaussian prior centered on the fitted value of a GP fit to the out-of-transit RV points, and the prior on the GP amplitude was controlled by a uniform (uninformative) prior between 0--10 m\,s$^{-1}$. These priors are also listed in Table~\ref{tab:prior}. 

\begin{table}
	\caption{The prior on free parameters in RM analysis.}
	\centering
	\begin{tabular}{lc}
    \hline \hline 
    	\noalign{\smallskip}
		Parameter & Prior \\
		\noalign{\smallskip}
		\hline
		\noalign{\smallskip}
        $\lambda$ (deg) & $\mathcal{U}(-180;180)$\\
        $v \sin{i_\star} $ $(km\,s^{-1})$&  $\mathcal{N}(7,3)$\\
        $T_{0} (day)$ & $\mathcal{N}(Ephem;0.01)$\\
         Linear limb darkening coefficient &  $\mathcal{N}(\texttt{LDTk};0.1)$\\
        $A_{GP} (m/s)$ &  $\mathcal{U}(0;10)$\\
        $\tau_{GP} (days)$ &  $\mathcal{N}(0.05;0.1)$\\
        \noalign{\smallskip}
		\hline
	\end{tabular}
	\label{tab:prior}
	\begin{flushleft} 
\textbf{Notes}: $\mathcal{U}(a;b)$ is a uniform prior with lower and upper limits of $a$ and $b$. $\mathcal{N}(\mu; \sigma)$ is a normal distribution with mean $\mu$ and width $\sigma$. Ephem corresponds to the predicted ephemerides.
\end{flushleft}
\end{table}

We randomly initiated the initial values for our free parameters for 30 MCMC chains inside the prior distributions. For each chain we used a burn-in phase of 500 steps, and then again sampled the chains for 5000 steps. Thus, the results concatenated to produce 150000 steps. We determined the best fitted values by calculating the median values of the posterior distributions for each parameters, based on the fact that the posterior distributions were Gaussian. The best fitted model of \texttt{PyAstronomy} and \texttt{PyAstronomy}+GP, and RM observations are shown in Figure~\ref{fig:PyastronomyGP}, and the posterior distributions are given in Figure~\ref{fig:cornerPyastronomyGP}. We report all the best fitted values in Table~\ref{Tab:parameters}.

The \texttt{PyAstronomy}+GP model results in a moderately better fit as indicated by a decrease in the RMS of the residual. We also performed a model comparison using Bayesian information criterion (BIC). We regard the difference between two models as significant
if $\Delta BIC > 5$ \citep{Liddle-07}. We found $\Delta BIC =6$ that supports the idea of fitting the observed RM with a GP given their noise. \footnote{We also did a model comparison between fitting the observation with only GP, and that led to a worse fit, with $\Delta BIC =10$ in favor of \texttt{PyAstronomy}+GP model.}

Overall, using \texttt{PyAstronomy}+GP analysis, we find TOI-1431b to be highly misaligned, with a projected obliquity $ \lambda = -155^{+20}_{-10}$ degrees.

\begin{figure}[h!]
  \includegraphics[width=0.5\textwidth]{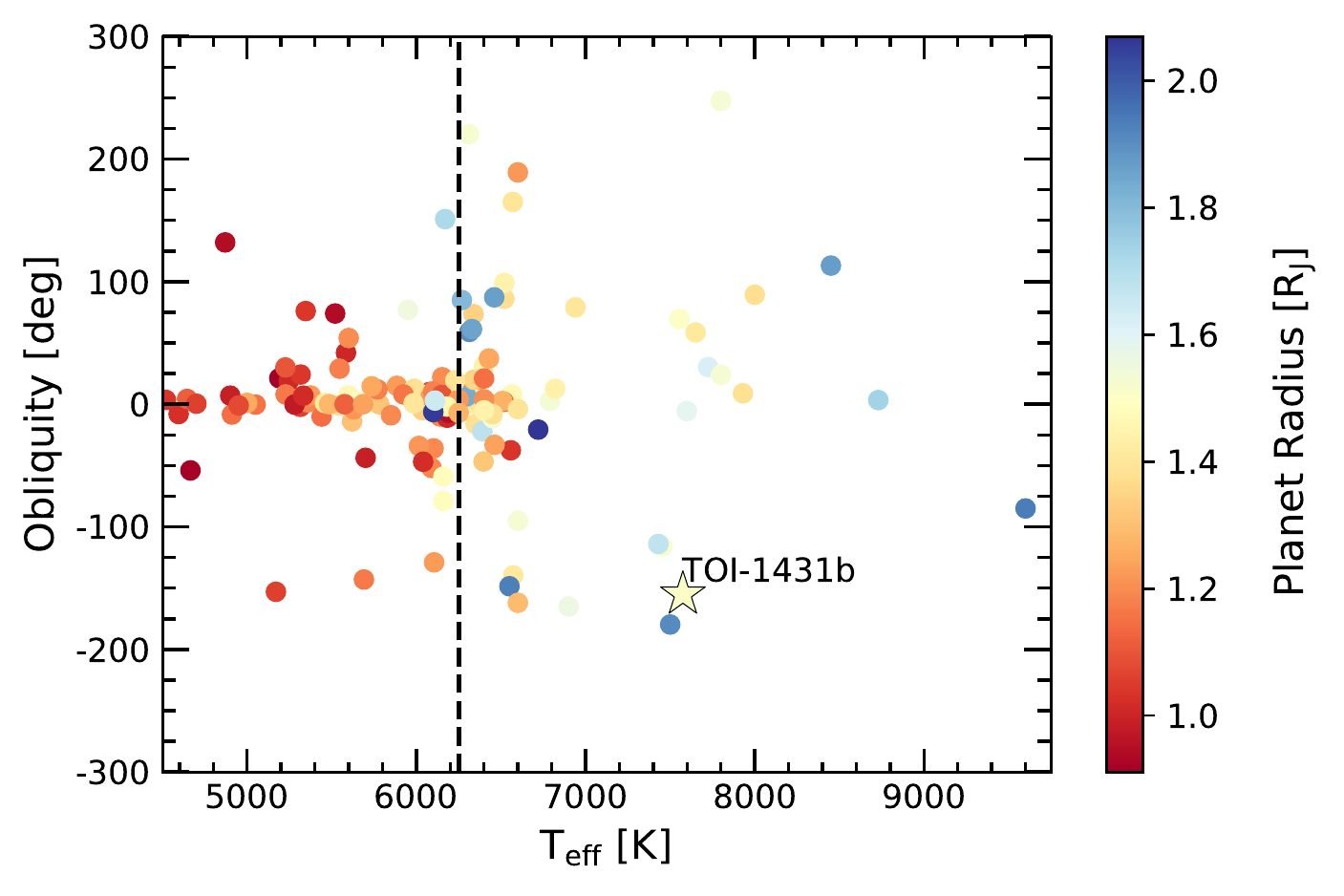}
  \caption{Distribution of measured orbital obliquity for the known transiting exoplanets and brown dwarfs as function of
their host star effective temperature (dots) and colour bar
presents the planetary radius, all extracted from TEPCat orbital obliquity catalogue \citep{TEPCat}. The star symbol represents TOI-1431b’s spin-orbit measurement. The black-dashed vertical line marks
the 6250 K effective temperature transition from \citet{Winn10}.}
  \label{fig:Obliquitymap}
\end{figure}

We show in Figure~\ref{fig:Obliquitymap} the distribution of measured projected obliquity for known transiting planets (from TEPCat orbital obliquity catalogue; \citealt{TEPCat}) as a function of their host star effective temperature. We also overplotted the TOI-1431b, which follows the general trend of planets orbiting stars with effective temperatures higher than ${\sim 6200}~{\rm K}$, which tend to be misaligned \citep{Winn10}.

\section{Atmospheric Cross-correlation Analysis}
\label{sec:cc}


In order to investigate the atmospheric composition of TOI-1431b, we used the cross-correlation technique to search for atomic and ionized species. In our analysis we used model transmission spectra of \ion{Fe}{I}, \ion{Fe}{II}, \ion{Ca}{II}, \ion{TiO}{} \citep{TiO_line_list, TiO_McKemmish_2019}, \ion{VO}{} (B. Plez, priv. comm.), VI, MgI, and NaI, generated using \texttt{petitRADTRANS} \citep{molliere_2019_petiRADTRANS}, which allows us to create high-resolution model templates for atoms and molecules at the typical temperatures of exoplanet atmospheres. To calculate the models, we assumed a surface gravity ($ \log{g_p}$) of 3.57, corresponding to a planetary mass of about 3.12$M_J$ (Addison et al., 2021 submitted). We also assumed solar abundance and an isothermal profile at the temperature of 4000 K. Following \citet{Hoeijmakers_2019_kelt9}, we set the absorption continuum at 1 mbar.

We corrected for Earth's telluric spectra using {\tt Molecfit} \citep{Molecfit1,Molecfit2} for HARPS-N spectra, and SELENITE \citep{2019_selenite} for the EXPRES spectra.
For outlier rejection and normalization we applied the methodology described in \citet{Stangret_2020_MASCARA-2}. For HARPS-N data, we divided the spectrum and models into 10000-pixels orders, due to computing limitations. For the EXPRES data, we used the spectrograph orders. We removed outliers by analysing the time evolution of each pixel, removing them when they deviate from fitted quadratic polynomial by more than $5 \sigma$. Additionally we corrected the reflex motion of the star as well as systemic velocities by fitting the linear polynomial to out-of-transit RVs. In a final step we divided each spectrum by the master-out spectrum, which was computed as the mean of all the out-of-transit spectra.

Using a radial velocity range of $\pm200~{\rm  km~s^{-1}}$ in steps of 0.8~km\,s$^{-1}$, we cross-correlated in the Earth's rest frame  each order with the models of  \ion{Fe}{I}, \ion{Fe}{II}, \ion{Ca}{II}, TiO, VO, \ion{V}{I}, \ion{Mg}{I}, and \ion{Na}{I}.

The Pearson cross-correlation coefficients $c(v,t)$ were calculated as:
\begin{equation}
c(v, t) = \frac{\sum_{i=0}x_i(t)T_i(v)}{\sqrt{\sum_{i=0}x_i(t)^{2}\sum_{i=0}T_i(t)^{2}}},
\end{equation}
where $x_i$ are the residuals in all orders at time $t$  and $T_i$ are the values of the template shifted to different velocities $v$ 

In the next step we shift the cross-correlation map to the planet rest frame using the formula for planet radial velocities $v_p$

\begin{equation}
\label{eq:vel}
    v_p(t, K_p)= K_p \sin{2 \pi \phi (t) + v_{bar}(t)}
,\end{equation}

where $K_p$ is the semi-amplitude of the planet radial velocity, $\phi (t)$ is the orbital phase of the planet, and $v_{bar}(t)$ is the barycentric velocity. 
Assuming that the $K_p$ value is unknown, we calculated the planet radial velocities ($v_p$) for range of $K_p$ values from 0 to 300~km\,s$^{-1}$, in steps of 1~km\,s$^{-1}$.

Excluding the ingress and egress data, we co-added the in-transit cross-correlation values for each $K_p$ value separately. To check the significance of the signal, we calculated its S/N for each $K_p$ value, following the same method as in \citet{birkby_2017_sysrem}, \citet{brogi_2018_kpmap}, \citet{alonso_floriano_2018_hd189733}, and \citet{sanchez_lopez_2019_hd209458}. We expect the planetary signal at 0~km\,s$^{-1}$ radial velocity and $K_p=188.15$ ~km\,s$^{-1}$. 

\begin{figure*}[h!]
  \includegraphics[width=\textwidth]{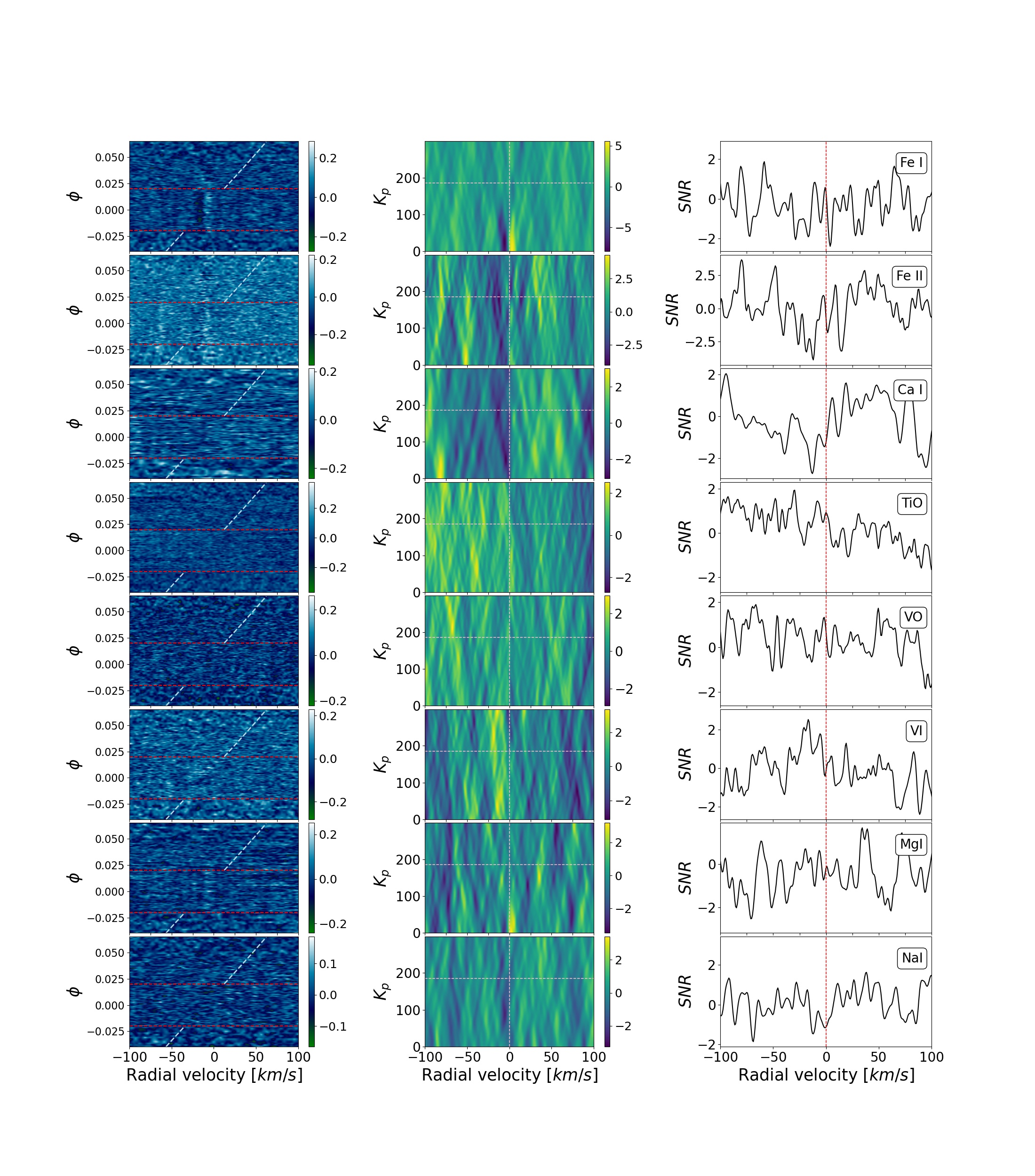}
  \caption{Left panels: Cross-correlation residual maps  of \ion{Fe}{I}, \ion{Fe}{II}, \ion{Ca}{II}, \ion{TiO}{}, \ion{VO}{}, \ion{V}{I}, \ion{Mg}{I}, and \ion{Na}{I} for combination of two night from HARPS-N. Red horizontal map shows beginning and end of the transit. Light-blue tilted line presents trace of expected signal from the planet. Middle panels: significance map for Kp in a range of 0 to 300~km\,s$^{-1}$, we expect signal from the planet in the 0~km\,s$^{-1}$ radial velocities and $K_p=185.3$ ~km\,s$^{-1}$ marked with dashed lines. Right panels: S/N plot for expected Kp value. Here, the RM+CLV effects have not been corrected.}  
  \label{fig:CC_molecules}
\end{figure*}

\begin{figure*}[h!]
  \includegraphics[width=\textwidth]{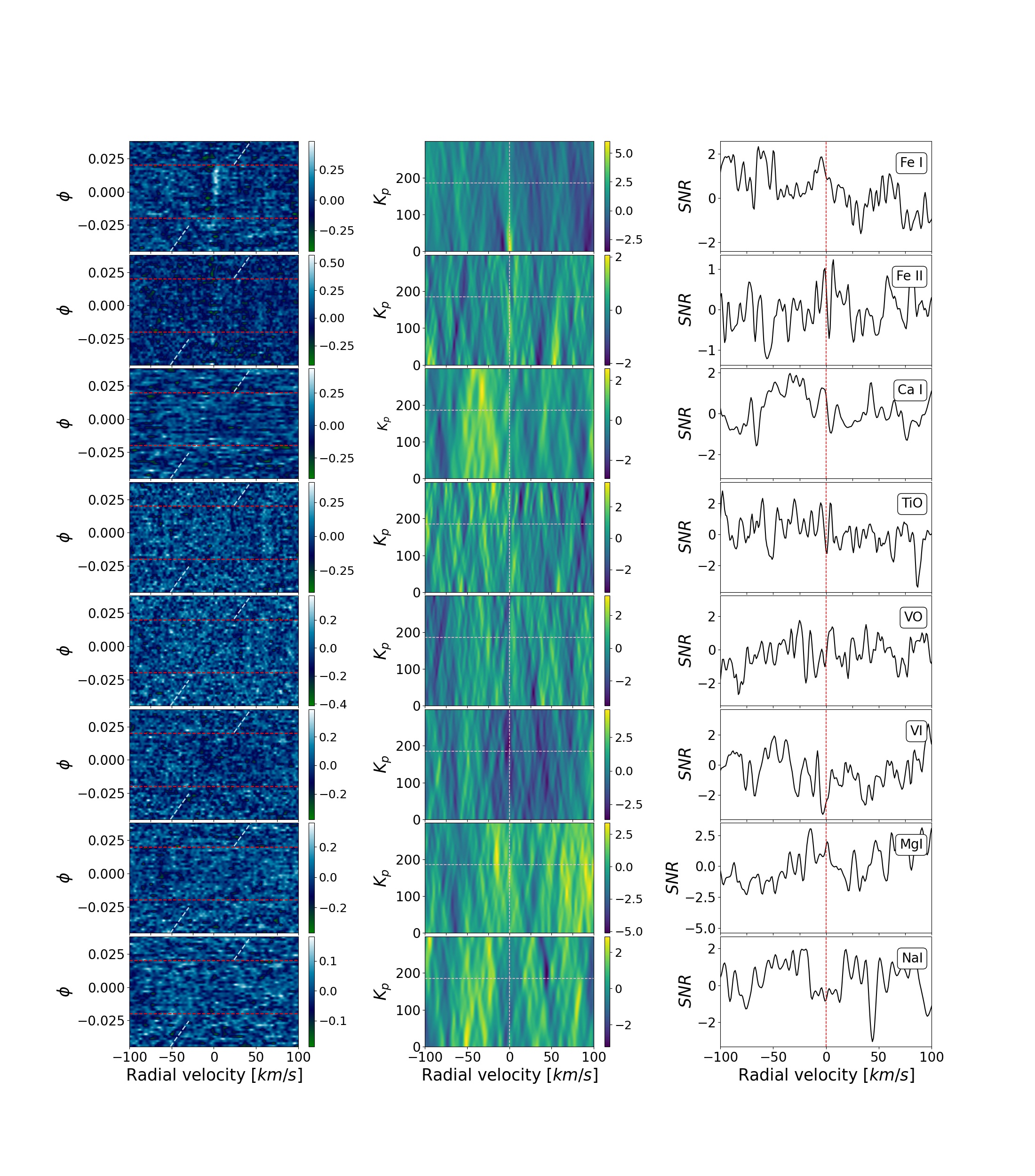}
  \caption{Same as Figure \ref{fig:CC_molecules}, but for EXPRES data.}
  \label{fig:CC_molecules_EXPRES}
\end{figure*}

In the left panels of Fig. \ref{fig:CC_molecules} and \ref{fig:CC_molecules_EXPRES} we show the cross-correlation maps for \ion{Fe}{I}, \ion{Fe}{II}, \ion{Ca}{II}, \ion{TiO}{}, \ion{VO}{}, VI, MgI and NaI for HARPS-N and EXPRES data, respectively. In the case of TiO,  we show the results using the line list by \citet{TiO_McKemmish_2019}, but we get very similar results using the line list by \citet{TiO_line_list}. The expected trace of the planetary signal is marked by a white tilted dashed line. In the middle panel we show a significance map for a range of $K_p$ values, from 0 to 300~km\,s$^{-1}$. The S/N was calculated by dividing the sum of in-transit cross-correlation residuals by the standard deviation of region from -50~km\,s$^{-1}$ to -150~km\,s$^{-1}$ and from 50~km\,s$^{-1}$ to 150~km\,s$^{-1}$, away from expected signal. The S/N plots at the expected $K_p$ value are given in the right panels. There is not significant planetary signal detection for any of the species investigated here.

TOI-1431 is a rotating star, and because of this, in the case of \ion{Fe}{I}, \ion{Fe}{II}, and \ion{Mg}{I}, we can clearly see the effects of the RM and center to limb variation (CLV) in the cross-correlation maps. These effects should be corrected as they might mask planetary absorption features. To this end, we modelled the stellar spectra at different planet orbital phases, which contains the RM and CLV effects. We applied the same methodology as described in \citet{Casasayas2019}. We used the \texttt{Spectroscopy Made Easy} tool (SME, \citet{SME}) to compute models using the Kurucz ATLAS9 and VALD3 line list \citep{VALD3}, and we modelled the stellar spectra for different limb-darkening angles. We assumed solar abundance and local thermodynamical equilibrium (LTE). Assuming also 1$R_P$=1.546 $R_J$ (Addison et al, 2021 submitted), we calculated the stellar models containing the RM and CLV effects taking into consideration the different stellar regions covered by the planet at each orbital phase. After dividing the model by the out-of-transit spectra we cross-correlated them with models of \ion{Fe}{I} and \ion{Fe}{II} using the same methodology applied for the observations. The same steps were taken to create the model considering only the RM effect.

\begin{figure*}[h!]
  \includegraphics[width=\textwidth]{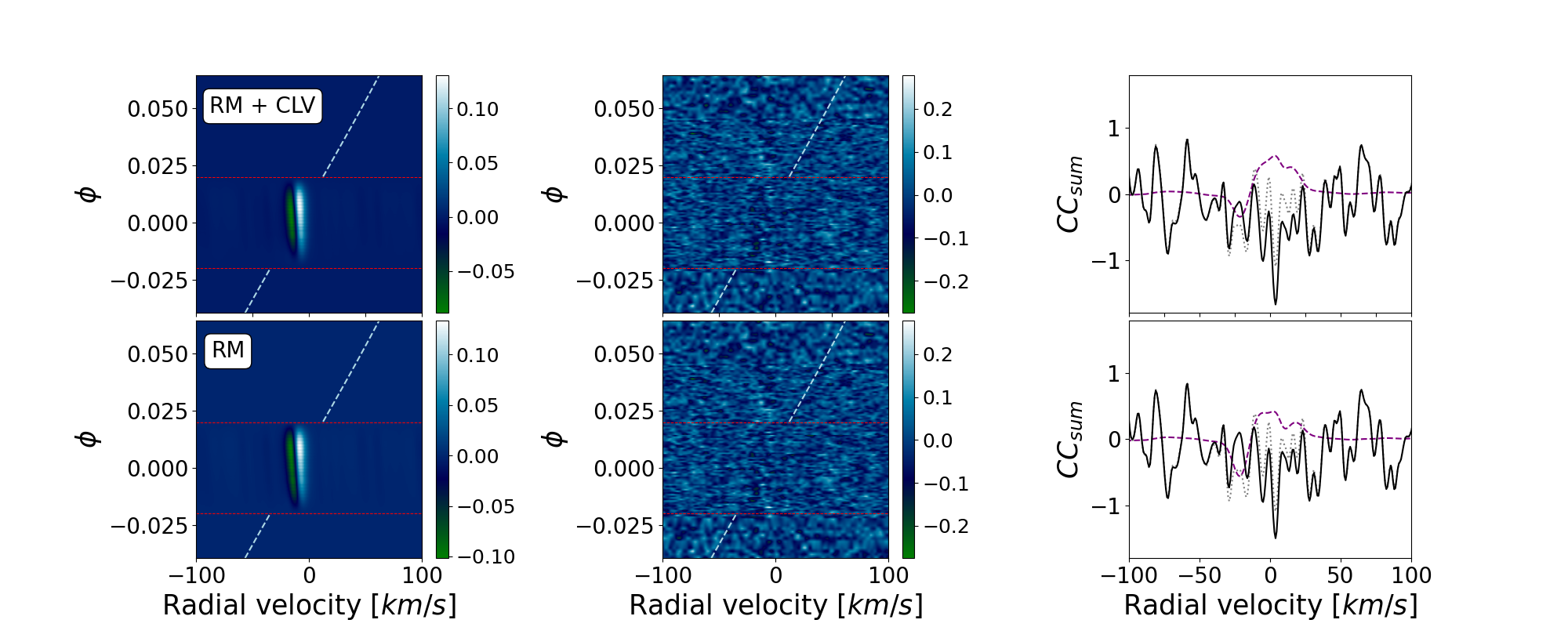} 
  \caption{Left panels: The RM plus CLV model (top) and the RM alone model (bottom) for \ion{Fe}{I}. Middle panels: cross-correlation residual maps of \ion{Fe}{I} for the combination of two HARPS-N nights after removing the modelled effects in the left panels. The red horizontal lines represent the beginning and end of the transit. The light-blue tilted dashed line shows the position of the expected signal of the planet. Right panels: cross-correlation sum plots at the expected $K_p$ value, before (gray dotted line) and after (black line) removing the effects. In purple we show the calculated models for each effect.}
  \label{fig:CC_FeI}
\end{figure*}

\begin{figure*}[h!]
\centering

     \includegraphics[width=1\textwidth]{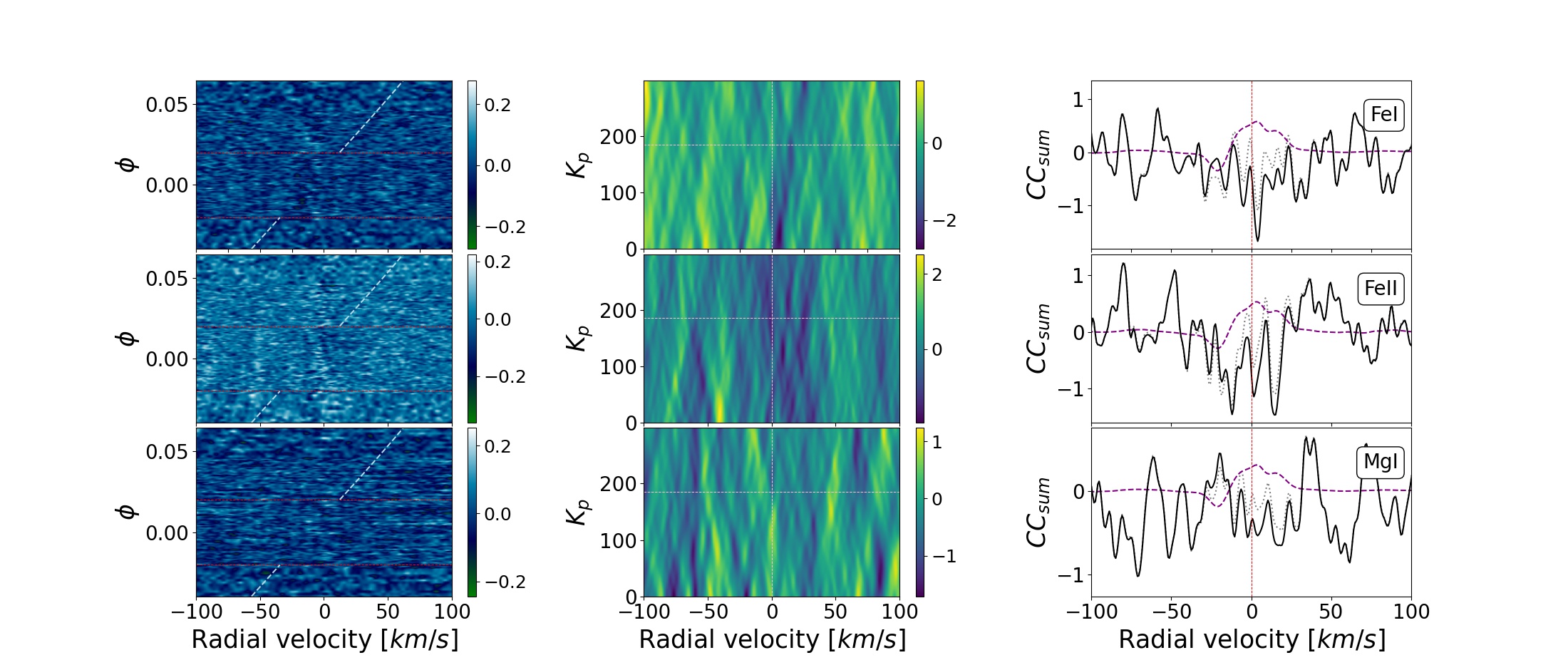}

  \caption{Left panels: cross-correlation residual maps for HARPS-N data after removing RM+CLV effects for Fe I (top), Fe II (middle), and MgI (bottom). Middle panels: Kp maps after removing RM+CLV signal. Right panels: cross-correlation sum plots at the expected $K_p$ value, before (gray dotted line) and after (black line) removing the effects. In purple we show the calculated models for RM+CLV effects.}
  \label{fig:FeI_FeI_MgI_HARPS_without_RM}
\end{figure*}

\begin{figure*}[h!]
\centering

     \includegraphics[width=1\textwidth]{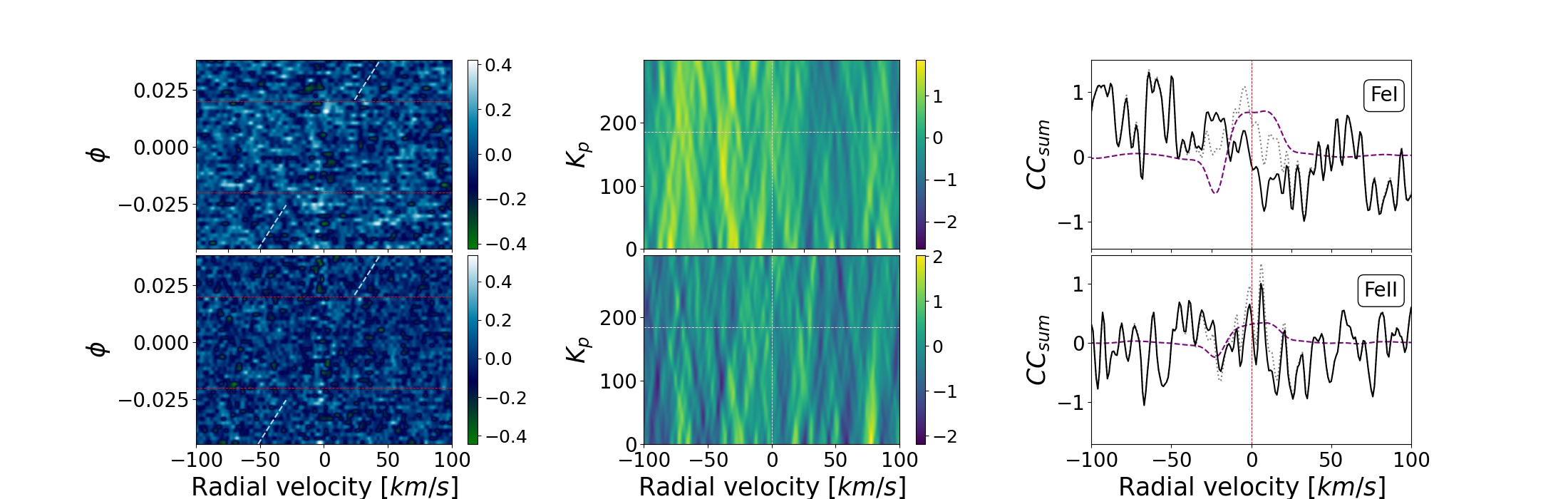}
  \caption{Same as Figure \ref{fig:FeI_FeI_MgI_HARPS_without_RM}, but for EXPRES data. In this case we do not show MgI, as no strong RM effect was detected in the EXPRESS data} 
  \label{fig:FeI_FeI_EXPRES_without_RM}
\end{figure*}

The next step was to remove the RM and CLV effects from the data. To do so we calculated the sum of all in-transit orbital phases for both data and models, and then scale the models to the data by fitting to the maximum value. Due to the uncertainties in the $\lambda$ value derived in Section~\ref{sec:rm} we observed small difference within the observed slope of the RM in the data and the models. In order to accurately remove the model effects, we shifted their slope to that measured in the observations. In Fig. ~\ref{fig:CC_FeI} we present the results after the RM+CLV and the RM alone effects correction for \ion{Fe}{I}. The models are presented in the left panels, while the middle panels present the cross-correlation residual maps after the corrections and the right panels show S/N plot at the expected Kp value before and after the correction. The corrections applied to the data seem to be quite efficient at removing the RM and CLV effects from the data, but do not lead to any significant detections. In figures \ref{fig:FeI_FeI_MgI_HARPS_without_RM} (for HARPS-N data) and \ref{fig:FeI_FeI_EXPRES_without_RM} (for EXPRES data), we present the results after removing the RM+CLV effects for the atomic species where the RM residuals were detected, namely (\ion{Fe}{I}, \ion{Fe}{II} and \ion{Mg}{I}). Any remaining feature in the cross-correlation residual maps are probably associated to stellar activity.

As a final check, we simply masked the region where RM and CLV effects appear strong in the models, and re-calculated the significance map as well as S/N plots with the unmasked data (not shown), but we do not detected any significant absorption signal.

We also explored in both cases the range of planet radial-velocity semi-amplitudes from $K_p = 300$ to $K_p =600$ km s$^{-1}$ (see section 5 discussion for details) without results.

\section{Transmission spectroscopy}

The atmospheres of UHJs are expected to be depleted of water and other molecular species \citep{Parmentier2018, helling_wasp_18}, due to their high equilibrium temperatures. However, the presence of H$\alpha$ absorption excess in the stellar spectrum has been previously observed in other UHJs, as a sign of atmospheric expansion and hydrogen escape \citep{YanKELT9}. Here we analyze the transmission spectrum of TOI-1431b at the H$\alpha$ line region.

\begin{figure*}[h!]
\centering
 \includegraphics[width=1\textwidth]{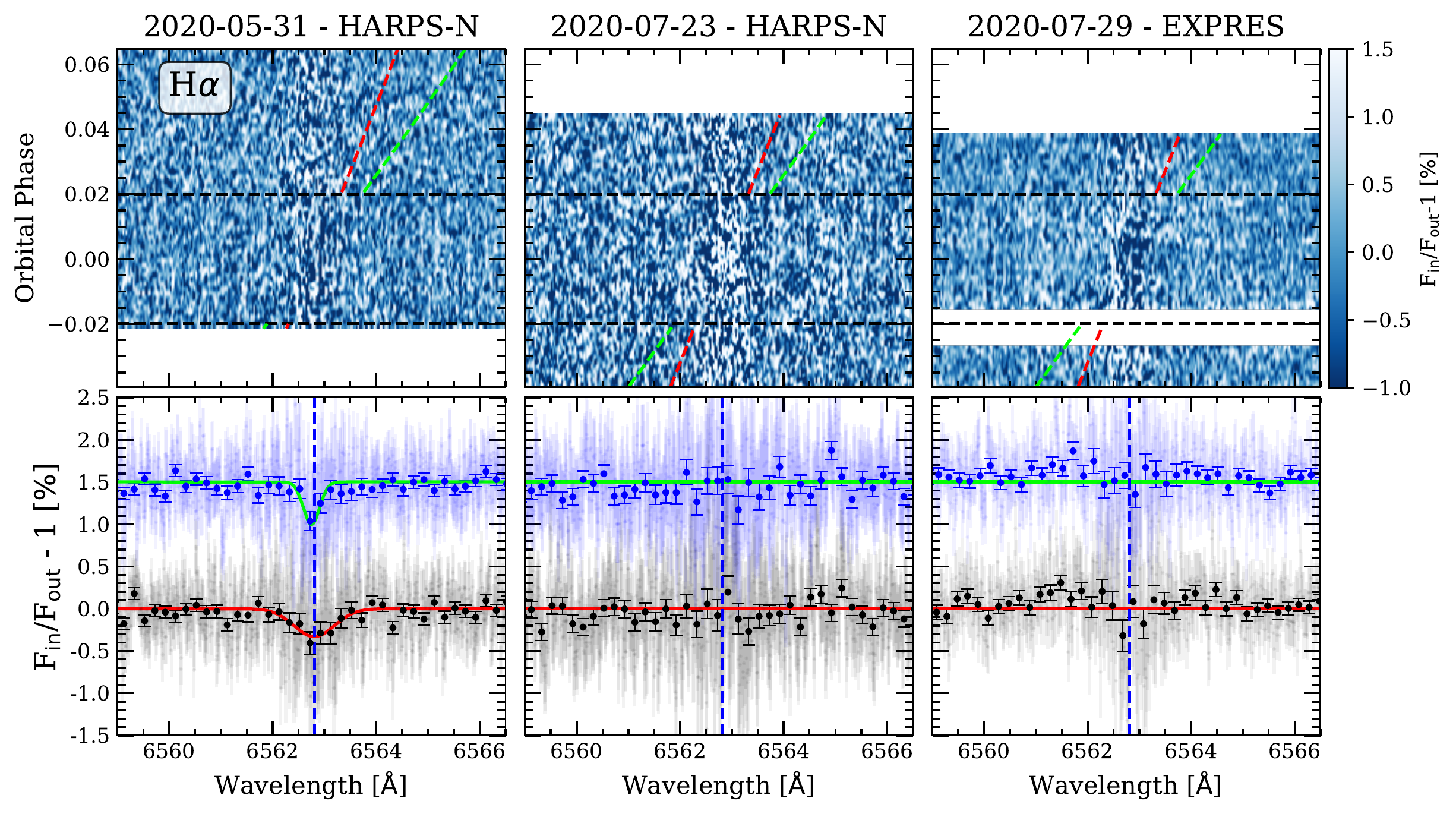}
  \caption{Top panels: Residual map around the H$\alpha$ line for the first night (left), second night (middle), and the EXPRES night (right). The horizontal black lines indicate the beginning and end of the transit. The red-dashed line shows the expected radial-velocity movement of the planet absorption ($K_p=188.15$\,km\,s$^{-1}$), plotted only in the out-of-transit regions for a better visualisation of the in-transit residuals. The green-dashed lines corresponds to $K_p=340$\,km\,s$^{-1}$. Bottom panels: Transmission spectrum of TOI-1431b around the H$\alpha$ line for Night 1 (left) and Night 2 (middle) obtained with the HARPS-N, and the night of 2020-07-29 (right) observed with EXPRES. In light gray we show the original data, and the black dots are the data binned at intervals of $0.2~{\rm \AA}$. In this case, the transmission spectrum is obtained assuming $K_p=188.15$\,km\,s$^{-1}$. In blue and with an offset of $1.5~\%$ for better visualisation, the transmission spectrum obtained at $K_p=340$\,km\,s$^{-1}$ is shown with the best fit Gaussian profile in green-dashed lines. The red line is the best fit Gaussian profile. The blue vertical line shows the laboratory position (at $6562.81~{\rm \AA}$) of the H$\alpha$ line. }
  \label{fig:Ha_comb}
\end{figure*}

The extraction of the transmission spectrum around individual lines is performed using the common methodology presented in \citet{Wytt2015} and \citet{Casasayas2019}. We first correct the observed stellar spectra of telluric absorption contamination from the Earth atmosphere using {\tt Molecfit} \citep{Molecfit1,Molecfit2}. Then, the spectra are shifted to the stellar rest frame using the barycentric Earth radial-velocity and the systemic velocity, and corrected by reflex motion of the star ($v_{bar}(t)$ Equation~\ref{eq:vel}, and $v_{sys}$ and reflex motion  retrieved from RVs), as discussed in Section~\ref{sec:cc}. Once all spectra are moved to the stellar rest frame, in order to remove the stellar contribution from the data, we divide each individual spectrum by the master out-of-transit spectrum, computed as the combination of all out-of-transit spectra. At this point, only variations of the stellar lines profile during the observations are expected to be observed in the residuals. Finally, the transmission spectrum is computed by combining all the resulting in-transit residuals once moved to the planetary rest frame using the weighted mean. The planet radial velocity during each exposure is calculated as presented in Section~\ref{sec:cc}. This combination is performed with a simple average of the relative flux per wavelength. The individual night results can be observed in Fig.~\ref{fig:Ha_comb}.

The features observed in the final transmission spectrum are quantified by fitting a Gaussian profile to the H$\alpha$ region for each individual night. The uncertainties of the transmission spectra come from the propagation of the photon noise and readout noise of the observations. These are then used to estimate the uncertainties of the best-fit values from the Gaussian profile by using the diagonal elements of the covariance matrix. For the Night 1 observed with HARPS-N we measure a $\sim 4.5\sigma$ absorption excess of $0.33\pm0.07~\%$ and a full width at half maximum (FWHM) of $0.9\pm0.2~{\rm \AA}$, with no significant blue shift ($0\pm4$\,km\,s$^{-1}$). For the Night 2 observed with the same instrument, no significant excess is observed in the transmission spectrum ($0.0\pm0.1\%$ with a $0.75~{\rm \AA}$ passband centred to the H$\alpha$ position). For the transit observed with EXPRES, the features observed at the H$\alpha$ position are consistent with null absorption excess at $1.4\sigma$, measuring an absorption excess of $0.13\pm0.09\%$ in a $0.75~{\rm \AA}$ passband. We note that the RM and CLV effects are not significant in transmission spectroscopy at the S/N achieved in the observations.

In Fig. \ref{fig:Ha_comb}, there is an absorption-like feature that appears only during the in-transit exposures for the first HARPS-N night (right panels in Fig. \ref{fig:Ha_comb}), but this absorption trail disappears for the second night (middle panels in Fig. \ref{fig:Ha_comb}) and for EXPRES observations (right panels in Fig. \ref{fig:Ha_comb}). 
Visually, it is difficult to recognize if the velocity of this absorption feature is in agreement with the predicted velocity of TOI~1431b or not. As presented in \citet{Chen2020}, in order to probe its velocity during the transit, we cross-correlate a Gaussian profile with each individual exposure of the 2D map. The Gaussian profile is computed with a contrast of $-0.4\%$ and FWHM of $0.75~{\rm \AA}$, slightly narrower and deeper than the best-fit obtained in the transmission spectrum for $K_p=188.15$\,km\,s$^{-1}$. We explore the velocity range between $\pm650$\,km\,s$^{-1}$ in steps of $0.5$\,km\,s$^{-1}$. Once the cross-correlation is applied to each exposure, we collapse the in-transit residuals using different $K_p$ values between $-100$ and $+700$\,km\,s$^{-1}$ in steps of 0.5\,km\,s$^{-1}$. The $K_p$-map is then computed as described in Section~\ref{sec:cc}. In this case, however, we use the range from -300 to -100\,km\,s$^{-1}$, and from +100 to 300\,km\,s$^{-1}$ to compute the standard deviation and calculate the S/N of the result. Although excess of absorption is observable at the predicted $K_p$, this excess is maximum at $K_p\sim340$\,km\,s$^{-1}$, far from the expected theoretical value (see Fig. \ref{fig:Ha_N1_CC}). The transmission spectrum computed considering this value is shown in the bottom panels of Figure~\ref{fig:Ha_comb}. For the Night 1, the transmission spectrum shows a contrast of $-0.50\pm0.09\%$ and FWHM of $0.4\pm0.1~{\rm \AA}$, shifted by $-3\pm1$\,km\,s$^{-1}$. However, there is no evidence of absorption during the other two nights using the same $K_p$ value.

Finally, we check the origin of these absorption features using the Empirical Monte Carlo (EMC) method \citep{Redfield2008}. The EMC is based on computing the transmission spectrum assuming different combinations of spectra building the in- and out-of-transit samples. Therefore, we would expect to reproduce the results only when the in- and out-of-transit exposures are correctly ordered. Here, we use the three common scenarios: 'in-in', 'out-out', and 'in-out' (see \citealt{Casasayas2019} for more details). We run the EMC $20~000$ times per scenario, and measure the absorption depth of every final transmission spectra using a $0.75~{\AA}$ ($\sim1$~FWHM) bandwidth centred on the expected H$\alpha$ position. The results of the individual nights are presented in Figure~\ref{fig:Ha_N1_EMC}. In all cases, we observe that the control samples ('in-in' and 'out-out') are centred at $\sim0~\%$ absorption depth, while the planet scenario ('in-out') is centred at $-0.17~\%$ for the Night 1 observed with HARPS-N , $+0.01~\%$ for the Night 2, and $-0.18~\%$ for the night observed with EXPRES. The standard deviation of the 'out-out' control distributions is $0.16~\%$, $0.24~\%$ and $0.24~\%$ for each night, respectively. This value is indicative of variations in the stellar lines core. Thus, we conclude that the most likely explanation for the variability observed in the H$\alpha$ absorption is stellar activity rather than variability in the planetary atmosphere.

We also explore the transmission spectrum of TOI-1431b around the \ion{Na}{i} doublet lines, but we find no excess of absorption for any of the individual nights(see Figure~\ref{fig:Na_comb}). 

\section{Discussion and Conclusions}

We observed the transmission spectrum of ultra-hot Jupiter TOI-1431b / MASCARA-5b during two nights using the HARPS-N high-resolution spectrograph and one night using the EXPRES spectrograph. Our results indicate that TOI-1431b does not seem to be the proto-typical UHJ.

By analysing the Rossiter–McLaughlin effect we find an obliquity value of $\lambda=-155$ deg. This puts the planet on a very inclined, near polar orbit, which speaks about an interesting  dynamical history, and perhaps indicating the presence of more than one planet in the early history for this system \citep{Triaud18}. The misalignment of TOI-1431b is also consistent with misaligned planets being preferentially found around stars with effective temperatures $T_{eff} > 6250$ K \citep{Albrecht2019}.

Additionally we studied the composition of TOI-1431b's atmosphere using two different methods. First, we used the cross-correlation technique to search for \ion{Fe}{I}, \ion{Fe}{II}, \ion{Ca}{I}, \ion{Na}{i}, \ion{Mg}{i}, \ion{V}{i}, \ion{TiO}{}, and \ion{VO}{}, finding no evidences of the presence of any of these atoms and molecules in its atmosphere. In the particular case of \ion{VO}{}, the inaccuracy of available line list remains a possible explanation for the non-detection. We also used transmission spectroscopy analysis to search for the $H\alpha$ absorption line and the \ion{Na}{i} doublet lines, again with negative results. These results are at odds with other studies of similar UHJs orbiting bright stars where various atoms and molecules have been found.

\begin{figure}[h!]
  \includegraphics[width=0.48\textwidth]{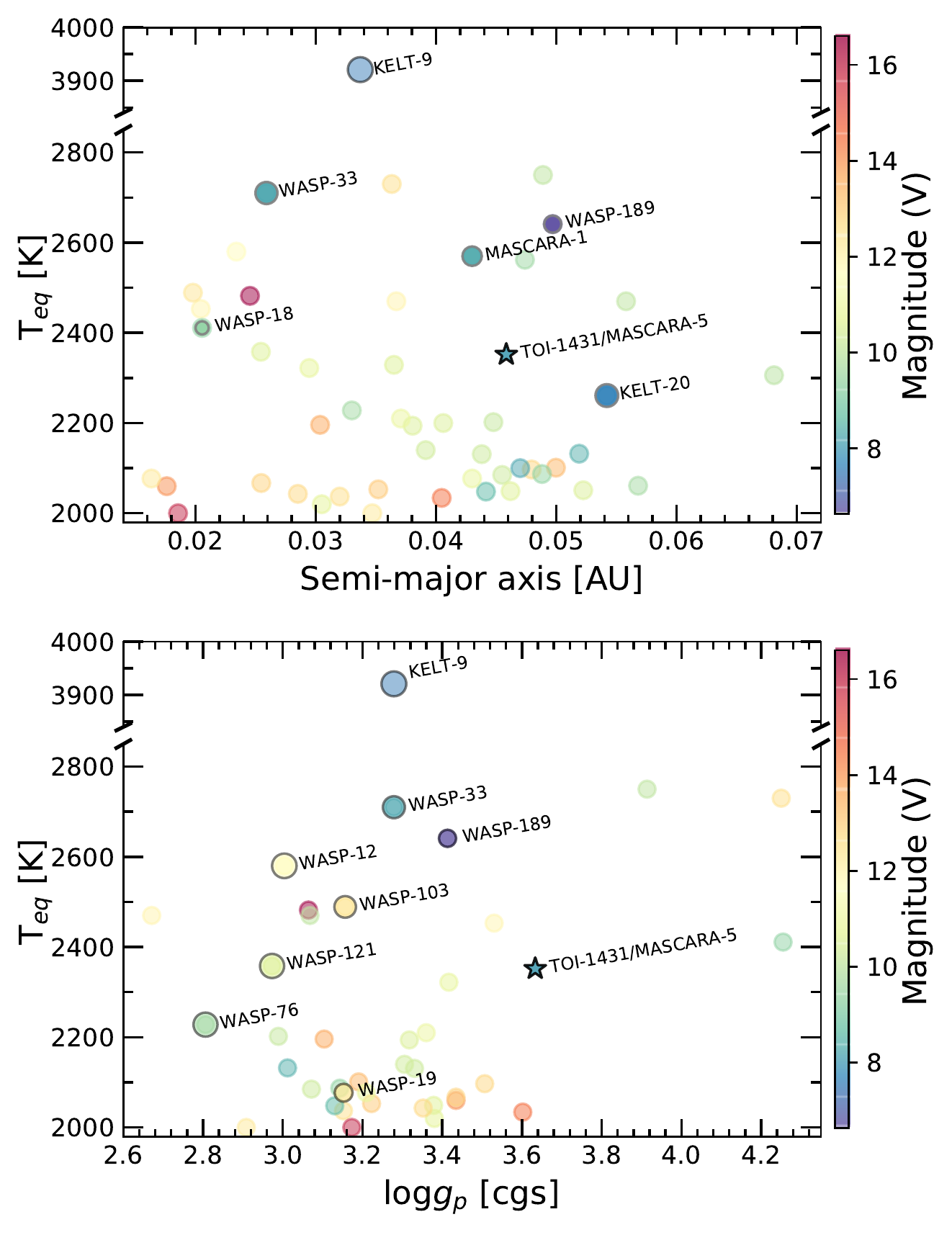}
  \caption{Top figure: Context of TOI-1431b, marked with a star symbol, with respect to all other known ultra hot Jupiter planets ($T_{eq}>2000$\,K). The planet-to-star distance (semi major axis) is shown in the horizontal axis, and the equilibrium temperature of the exoplanets is shown in the vertical axis. We note that only those planets with $R_p>0.6~R_J$ are shown. Bottom figure: Context of TOI-1431b, marked with a star symbol, with respect to all the UHJ with known surface gravity (we excluded all planets whose masses have only upper limits). The equilibrium temperature is shown in the vertical axis and surface gravity of the planets in the horizontal axis. The marked planets are planets which atmospheres were studied before. For both figures the V band magnitude of the host star is colour-coded and the markers size is indicative of planet's radius.  The data is extracted from TEPCat catalogue \citep{TEPCat}. }
  \label{fig:UHJ_sample}
\end{figure}

Figure~\ref{fig:UHJ_sample} puts TOI-1431b in context by presenting all known UHJs with their equilibrium temperature plotted against semi-major axis.  
The absence of ionized atomic species, most specially \ion{Fe}{II}, in the atmosphere of TOI-1431b poses a mystery, as this ionized Fe has been detected in almost all UHJs for which precise high-resolution spectroscopic data are available, including planets with lower $T_{eq}$ than TOI-1431b. For example, MASCARA-2b/KELT-20b has a $T_{eq}$ value about 100 K lower, and \ion{Fe}{II} has been detected in its atmosphere both in cross-correlation \citep{Stangret_2020_MASCARA-2,Nugroho2020_KELT20,Hoeijmakers_mascara2} and in transmission spectroscopy for the strongest individual lines \citep{Casasayas2019}. 

A first reason for the absence of atmospheric features could be the S/N of the observations. However, TOI-1431 is nearly as bright as MASCARA-2 or KELT-20 (only 0.4 mag fainter in V), and in both cases the same instrumentation has been successfully used to retrieve significant detections \citep{Hoeijmakers_2019_kelt9, Casasayas2019}. Low S/N could be perhaps the culprit for the non-detection of H$\alpha$ absorption during the second night observed with HARPS-N and the observations performed with EXPRES, but can hardly explain the negative cross-correlation results during the first night. We also performed some injection tests (not shown here) where we were able to recover from the data planetary signals with strengths similar to those detected in other UHJs, reinforcing our hypothesis that S/N is not the limiting factor for atmospheric signatures detection.

The second possibility is the influence of the Rossiter–McLaughlin effect on the data, being TOI-1431 on a nearly polar orbit. \citet{Casasayas2020} already demonstrated how inappropriately dealing with this effect can lead to spurious detections or mask the planetary signals. Although these effects are clearly seen in the cross-correlation residual maps of Figure~\ref{fig:CC_molecules}, the exoplanet atmospheric features, if present, could also be visually disentangled in the regions where the planet radial-velocity and RM effect do not overlap. As the exoplanet atmosphere is not clearly seen in the maps, we attempted to correct both the RM and the CLV effect on the data, by modeling the effects and subtracting them from the observations, but still no significant absorption is found. Nevertheless, the impact of these effects in the planet rest frame remains in the noise level of the data, and thus are not a likely explanation for the non-detections.

The third and most probable explanation lies in the bulk properties of the planet itself. TOI-1431b has a large surface gravity and a small scale height, similar to that of the UHJ WASP-189 b for which no consistent atmospheric absorption have been detected \citep{wasp-189-2020-no_detection}. These two planets have the highest surface gravity among the UHJ planets whose atmospheres have studied so far. If this is the case, observing the planetary emission spectrum, rather than the transmission spectrum, might give much more insight into the physics of these more massive UHJs planets. New higher resolution data with larger aperture telescopes might be also able to shed new light into this interesting UHJ planet.

%

\begin{acknowledgements}
 Based on observations made with the Italian Telescopio Nazionale Galileo (TNG) operated on the island of La Palma by the Fundación Galileo Galilei of the INAF (Istituto Nazionale di Astrofisica) at the Spanish Observatorio del Roque de los Muchachos of the Instituto de Astrofisica de Canarias.
 This work is partly financed by the Spanish Ministry of Economics and Competitiveness through project PGC2018-098153-B-C31.
 M. S. and N.C.B acknowledge the support of the Instituto de Astrofísica de Canarias via an Astrophysicist Resident fellowship.
 F.Y. acknowledges the support of the DFG priority program SPP 1992 "Exploring the Diversity of Extrasolar Planets (RE 1664/16-1)".
 
 This work made use of PyAstronomy and of the VALD database, operated at Uppsala University, the Institute of Astronomy RAS in Moscow, and the University of Vienna.

\end{acknowledgements}

\bibliographystyle{m2.bst} 
\bibliography{m2.bib} 
  

\onecolumn

\begin{appendix}

\section{Additional figures}
\label{sec:ap_indiv}

\begin{figure}[h!]
  \includegraphics[width=\textwidth]{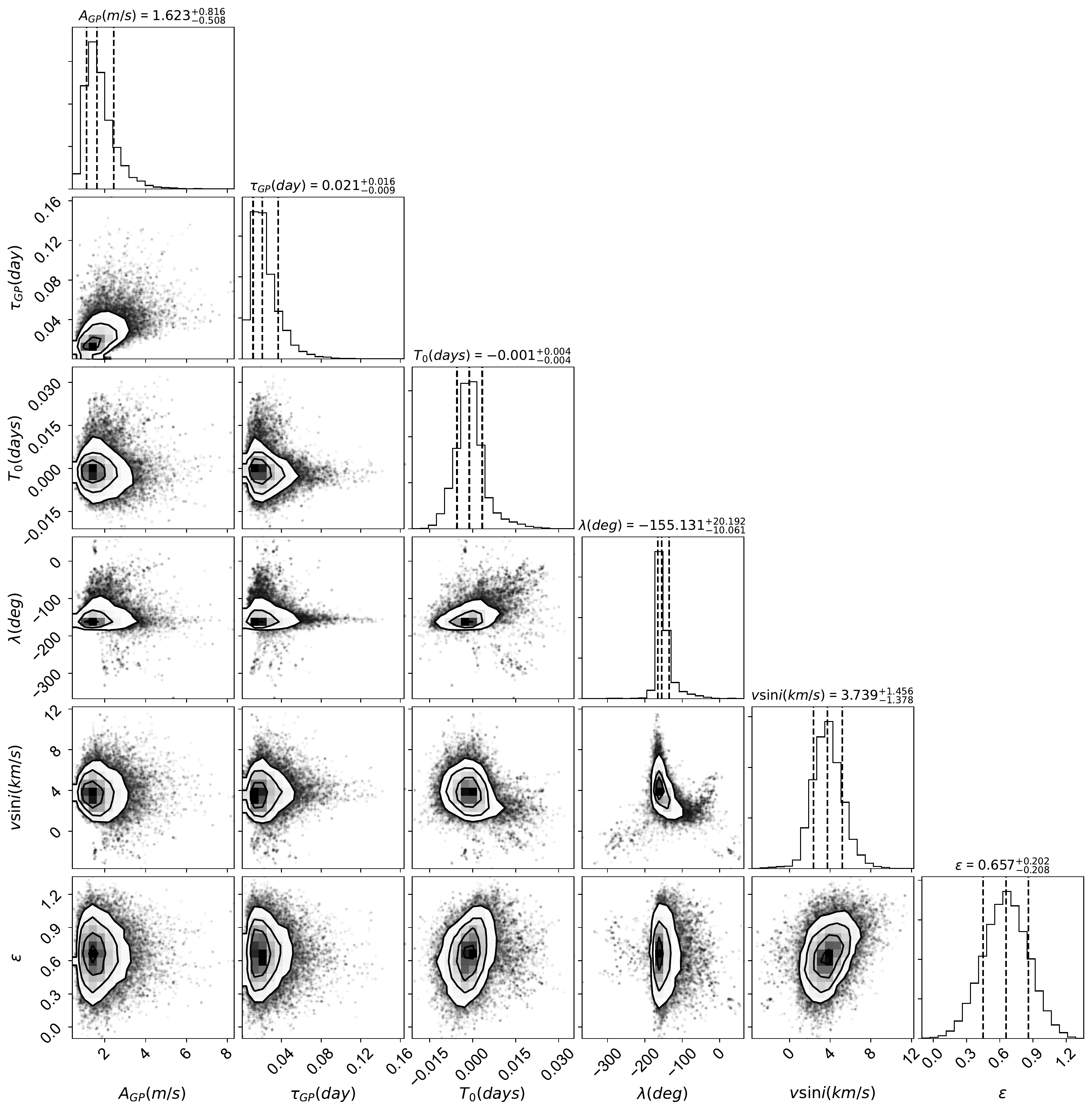}
  \caption{Best fit of Pyastrononmy+GP}
  \label{fig:cornerPyastronomyGP}
\end{figure}

\begin{figure*}[h!]
\centering
  \includegraphics[width=0.45\textwidth]{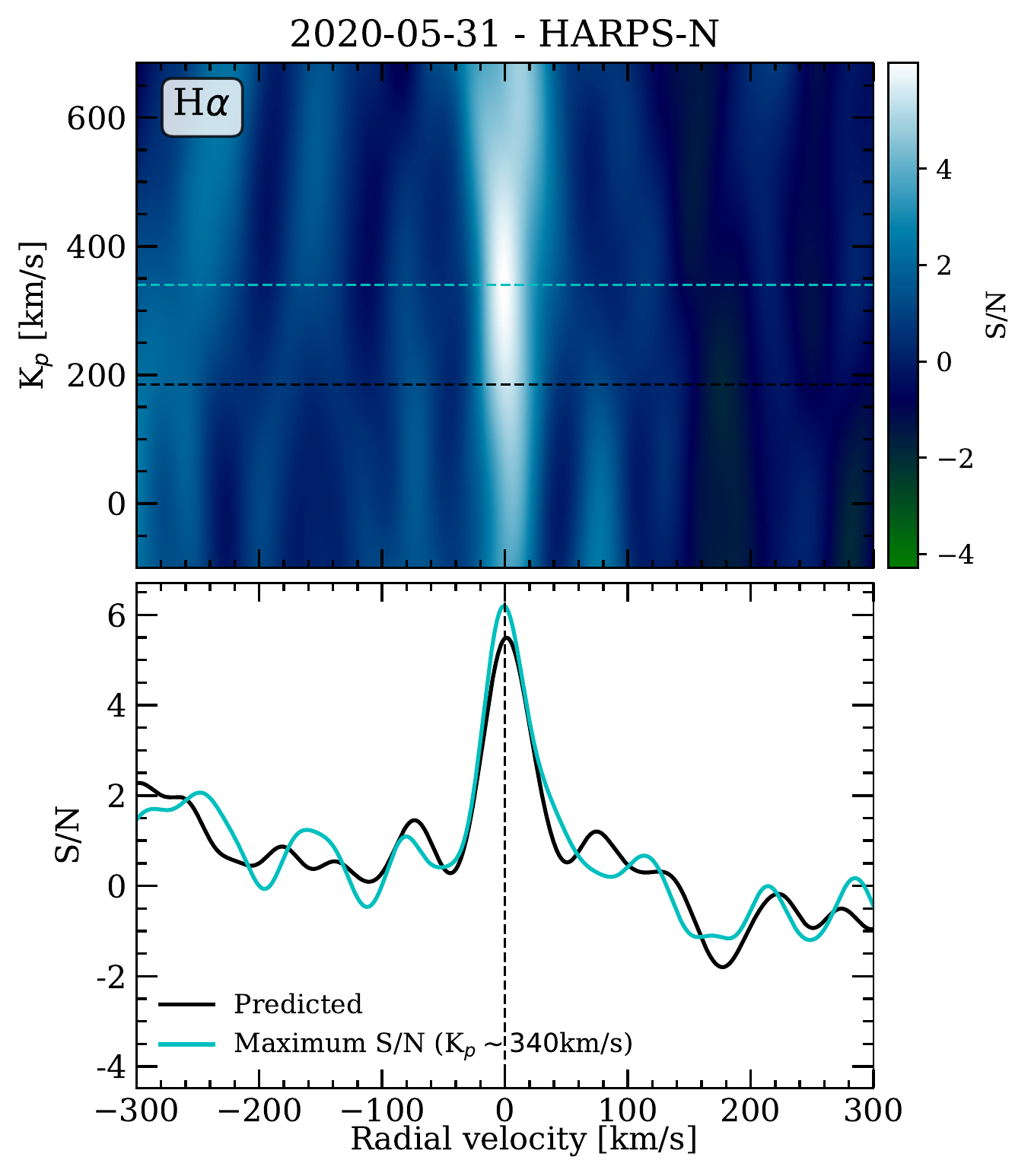}
  \caption{Cross-correlation results around the H$\alpha$ line of the first night. \textit{Top panel:} K$_p$ map of the cross-correlation values. The cyan-dashed line indicates the K$_p$ position with maximum S/N. The horizontal black dashed line shows the predicted K$_p = 188.15$\,km\,s$^{-1}$ value, the cyan-dashed line the K$_p \sim 340$\,km\,s$^{-1}$ at which the S/N is maximum, and the vertical black dashed line shows $0$\,km\,s$^{-1}$ shift. \textit{Bottom panel:} cross-correlation values extracted at the maximum S/N K$_p$ (cyan), and at the predicted K$_p$ (black).}
  \label{fig:Ha_N1_CC}
\end{figure*} 

\begin{figure*}[h!]
\centering
  \includegraphics[width=0.32\textwidth]{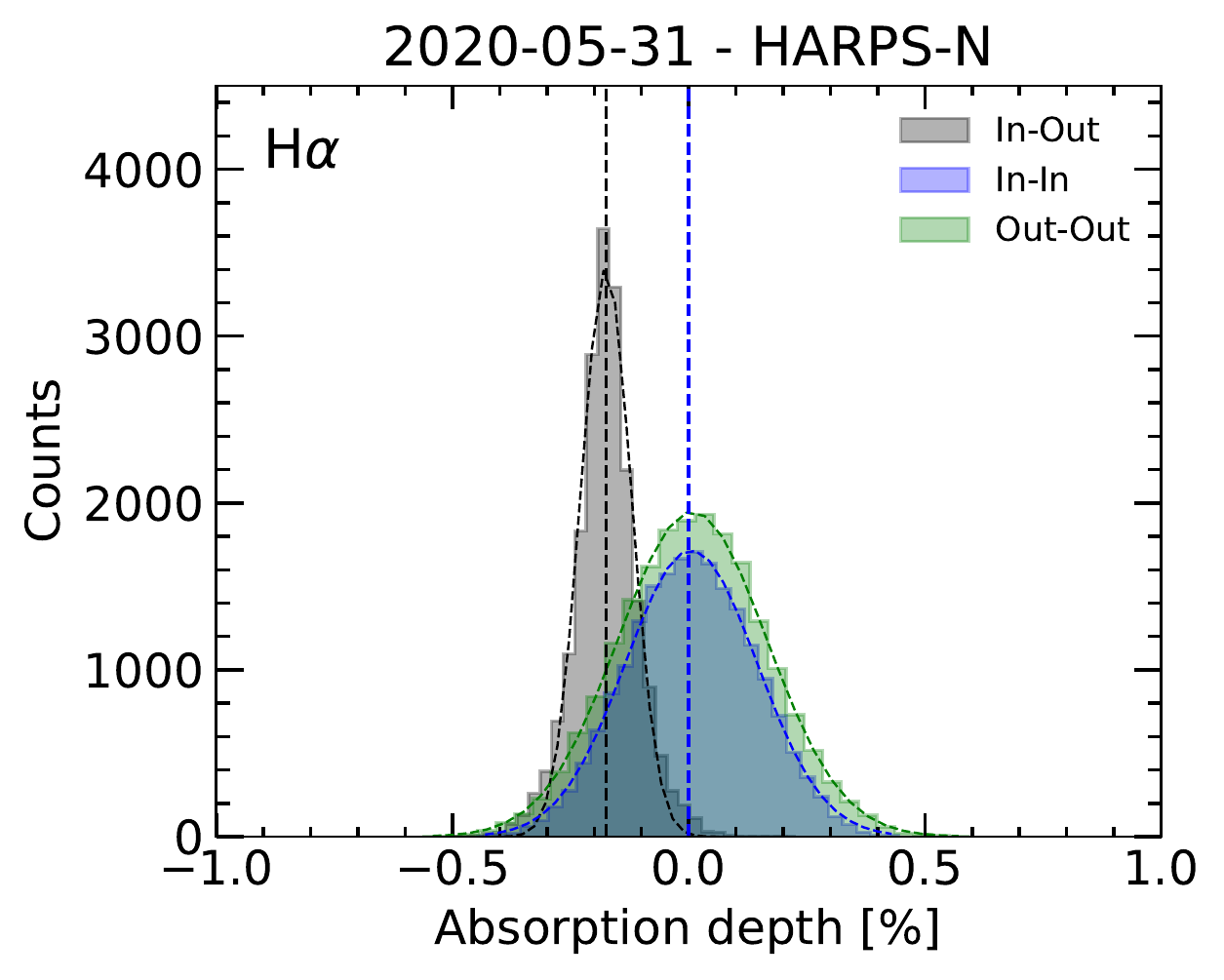}
   \includegraphics[width=0.32\textwidth]{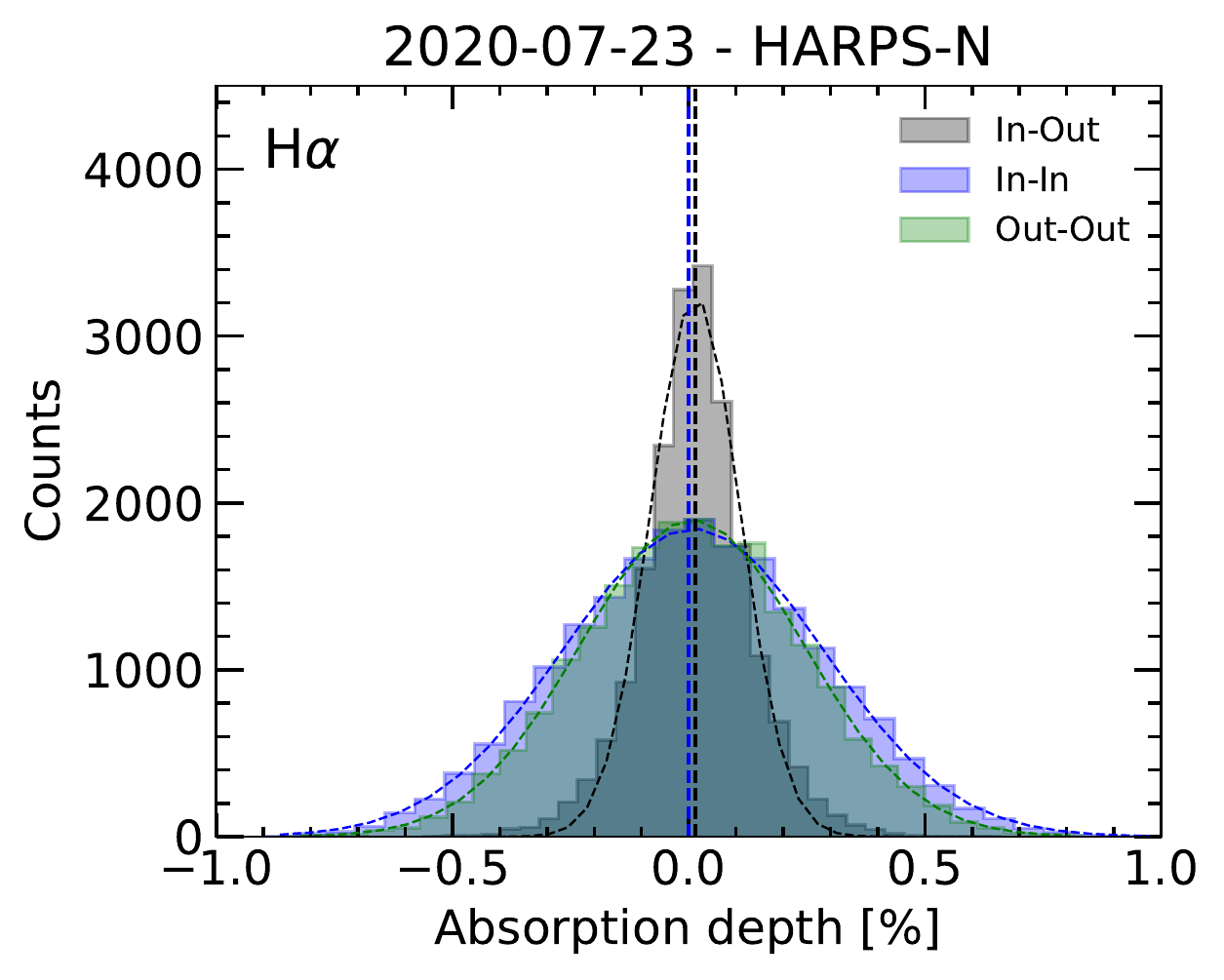}
     \includegraphics[width=0.32\textwidth]{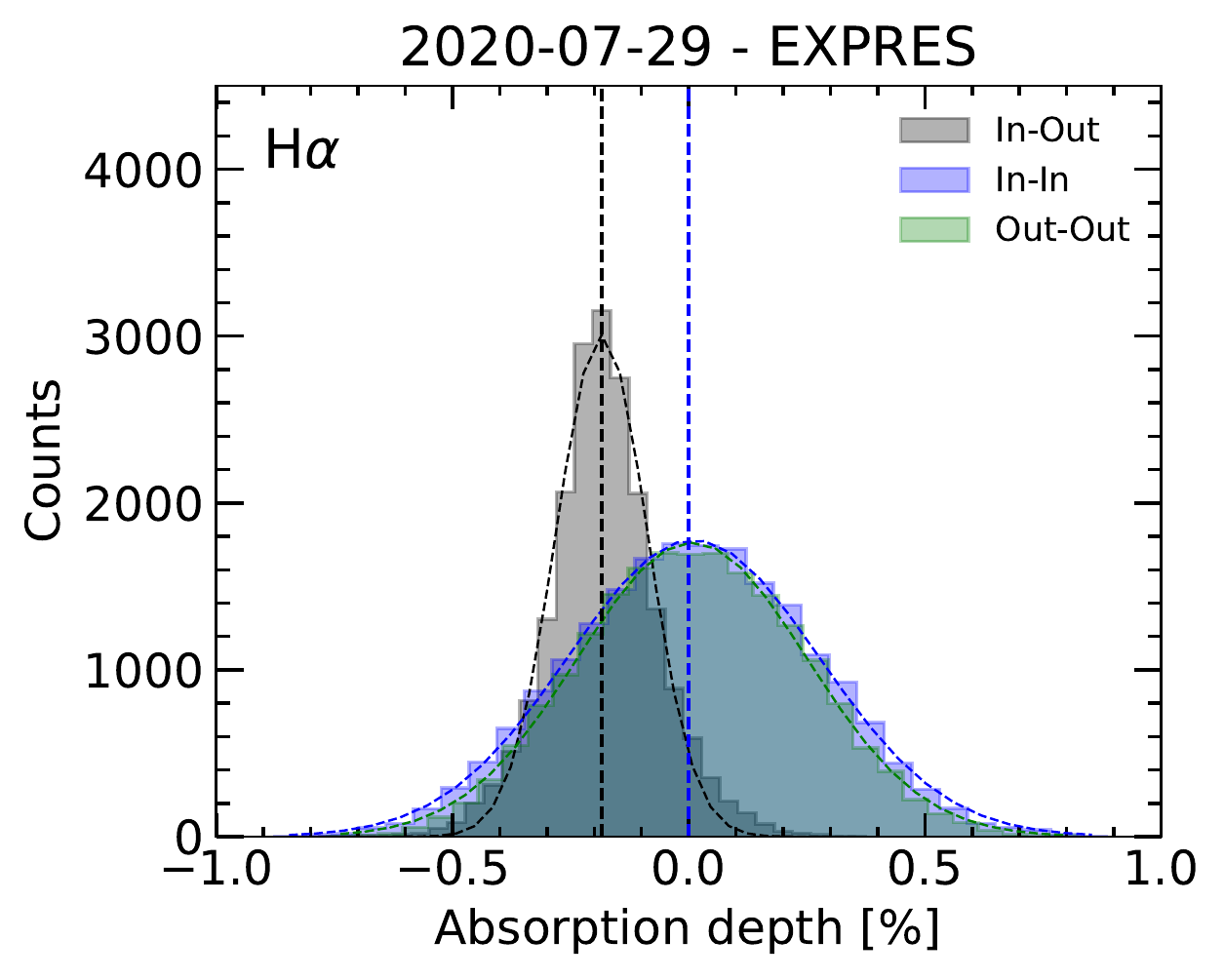}
  \caption{Empirical Monte Carlo (EMC) distributions in the H$\alpha$ line for the first (left) and second (middle) nights observed with HARPS-N, and the night observed with EXPRES (right). The distributions are obtained using 20~000 iterations and measuring the absorption depth with a bandwidth of $0.75~\AA$. Each panel corresponds to the analysis of one night. In green we present the `out-out' scenario, in blue the `in-in', and in grey the `in-out', which corresponds to the atmospheric absorption scenario. The blue-dashed vertical line marks the zero absorption level and the black-dashed line the center of the 
 `in-out’ distribution.}
  \label{fig:Ha_N1_EMC}
\end{figure*} 

\begin{figure*}[h!]
\centering
  \includegraphics[width=1\textwidth]{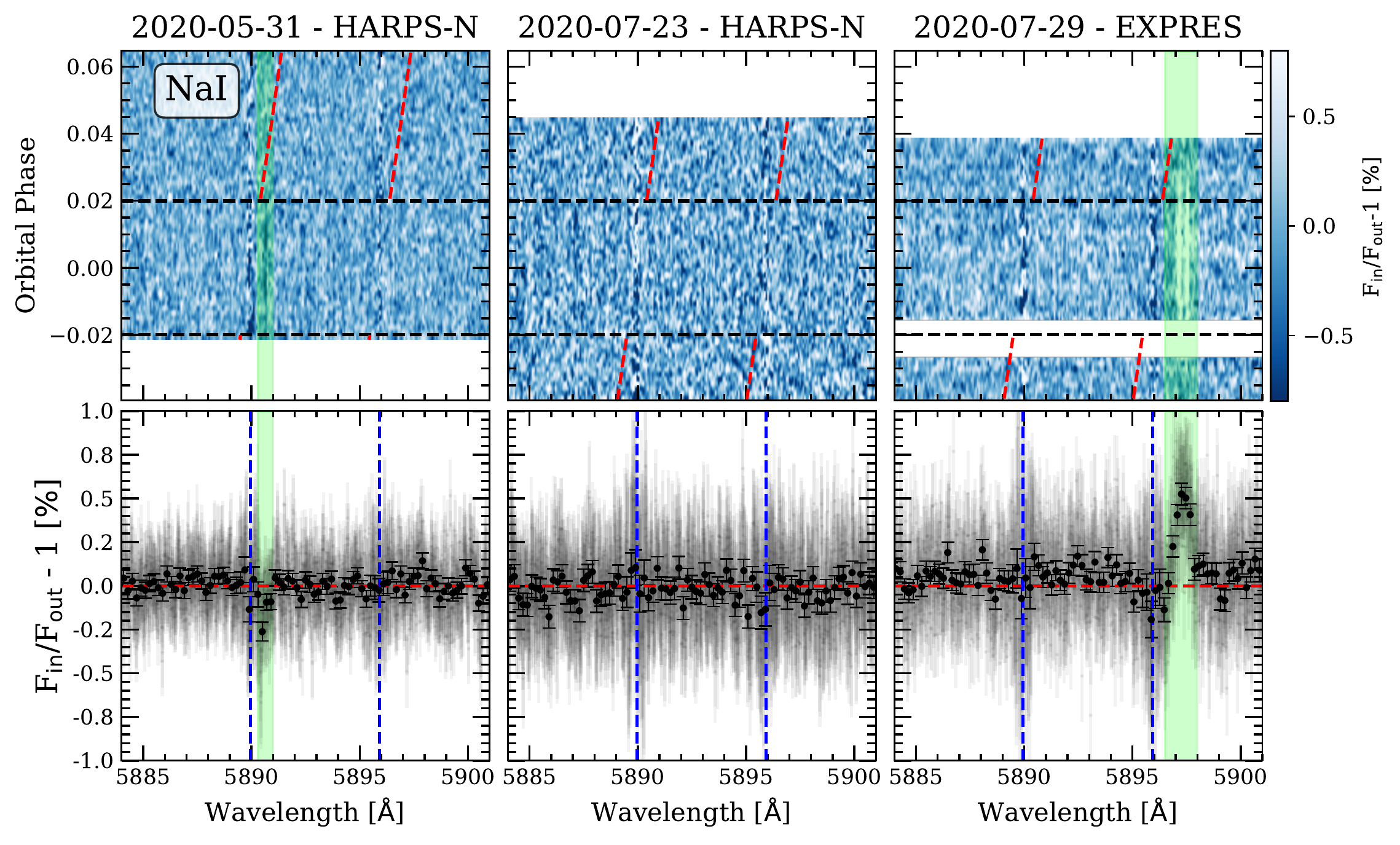}
  \caption{Same as Figure~\ref{fig:Ha_comb} but for the \ion{Na}{i} doublet. In light green we mark the regions affected by telluric residuals. The residual observed in the left panel (2020-05-31) corresponds to a telluric \ion{Na}{i} emission line which has not been completely corrected in the sky subtraction, probably due to the different efficiency of the two fibers. The residual observed in the right panel (2020-07-29) is produced by two H$_2$O absorption lines that are not detected by SELENITE in the telluric correction.}
  \label{fig:Na_comb}
\end{figure*}

\end{appendix}

\end{document}